\newif\ifanonym
\anonymtrue 

\documentclass[preprint,12pt,authoryear]{elsarticle}

\usepackage{amsmath,amssymb,amsfonts}
\usepackage{graphicx}
\usepackage{xcolor}

\usepackage{booktabs}
\usepackage{multirow}
\usepackage{subcaption}

\usepackage[hidelinks]{hyperref}
\usepackage[noabbrev,nameinlink]{cleveref}

\usepackage{enumitem}
\setlist{nosep}

\usepackage{amsmath,amssymb,amsfonts}
\usepackage{algorithmic}
\usepackage{graphicx}
\usepackage{dblfloatfix}

\usepackage{multirow}
\usepackage{tabularx}
\usepackage{booktabs}
\usepackage{enumitem}
\usepackage{longtable}
\usepackage{array}

\newdimen\xfigwd \xfigwd=0pt

\usepackage{adjustbox}

\journal{Ethics and Information Technology}

\begin{document}

\begin{frontmatter}

\title{From Review to Design: Ethical Multimodal Driver Monitoring Systems for Risk Mitigation, Incident Response, and Accountability in Automated Vehicles}

\ifanonym
% ---- Double-blind / anonymous submission ----
% \author{Anonymous Author(s)}
% Avoid institution names, emails, and corresponding-author markers in blind review.
% \address{Anonymous affiliation(s)}

% \else

\author[1]{Bilal Khan\corref{cor1}}
\ead{bilal.khan@universityofgalway.ie}

\author[1]{Waseem Shariff}
% \ead{waseem.shariff@universityofgalway.ie}

\author[2]{Rory Coyne}
% \ead{rorycoyne@rcsi.com}

\author[1]{Muhammad Ali Farooq}
% \ead{muhammadali.farooq@universityofgalway.ie}

\author[1]{Peter Corcoran}
\ead{peter.corcoran@universityofgalway.ie}

\cortext[cor1]{Corresponding author}

% \address[1]{School of Engineering, University of Galway, Galway, Ireland}
\address[1]{C3I Imaging Lab, School of Engineering, University of Galway}
\address[2]{Royal College of Surgeons in Ireland }

\fi

\begin{abstract}
As vehicles transition toward higher levels of automation, Driver Monitoring Systems (DMS) have become essential for ensuring human oversight, safety, and regulatory compliance in a vehicle. These systems rely on multimodal sensing and AI-driven inference to assess driver attention, cognitive state, and readiness to take control. While technologically promising, their deployment introduces a complex set of ethical and legal challenges - ranging from privacy and consent to data ownership and algorithmic fairness. While overarching frameworks such as the GDPR, EU AI Act, and IEEE standards offer important guidance, they lack the specificity required for addressing the unique risks posed by in-cabin sensing technologies. 

This paper adopts a review-to-design perspective, critically examining existing regulatory instruments and ethical frameworks -- such as the GDPR, the EU AI Act, and IEEE guidelines -- and identifying gaps in their applicability to the distinctive risks posed by multimodal, AI-enabled in-cabin monitoring. Building on this review, we propose a modular ethical design framework tailored specifically to Driver Monitoring Systems. The framework translates high-level principles into actionable design and deployment guidance, including user-configurable consent mechanisms, fairness-aware model development, transparency and explainability tools, and safeguards for driver emotional well-being.

Finally, the paper outlines a risk analysis and failure mitigation strategy, emphasizing proactive incident response and accountability mechanisms tailored to the DMS context. Together, these contributions aim to inform the development of transparent, trustworthy, and human-centered driver monitoring systems for next-generation autonomous vehicles.
\end{abstract}


\end{frontmatter}

\section{Introduction}
\label{sec:introduction}

% \subsection{Background Technology Landscape}
Driver Monitoring Systems (DMS) are increasingly recognized as critical components in the safe deployment of automated driving systems. As artificial intelligence (AI) accelerates innovation in the automotive industry, autonomous vehicles (AVs) have emerged at the forefront of this transformation \citep{garikapati2024autonomous}. However, despite ongoing advances toward higher levels of automation - particularly Levels 2 to 4 as defined by the Society of Automotive Engineers (SAE) - human oversight remains essential for ensuring safety, reliability, and public trust \citep{cunningham2015autonomous, sae2021sae}. 

Ethical concerns in AVs have received growing attention, especially in relation to uncertainty in decision-making, liability during system failures, and the delegation of responsibility between human and machine. In a study on Tesla’s Autopilot, the authors argued that ethical challenges such as decision-making capacity and delayed human intervention could be partially mitigated through the integration of advanced DMS \citep{jatavallabha2024tesla}. Another study emphasized that DMS should work in concert with AV systems to improve human-machine interaction and address both ethical and operational limitations in automated driving \citep{coyne2024understanding}. 

DMS technologies use in-cabin cameras, eye-tracking, and biometric sensors to assess a driver's cognitive and physical states in real time \citep{hayley2021driver}. These systems detect fatigue, distraction, or medical emergencies and contribute to accident prevention and road safety. Given the ubiquity of driving as a mode of transport, incorporating health-monitoring features into vehicles represents a scalable solution for enhancing public safety \citep{visconti2025innovative}. However, the very features that make DMS effective - continuous sensing and data collection - raise significant ethical challenges. Critics have likened these systems to Orwellian surveillance tools, with phrases such as "Big Other is Watching You" reflecting public discomfort with privacy loss and diminished autonomy \citep{zuboff2015big, gruchmann2025big}. Professional drivers may also experience psychological strain from constant monitoring, compounding workplace stress and resistance to adoption \citep{gruchmann2025big, bhoopalam2023long}. 

% \subsection{Background - Ethical and Regulatory Challenges}
Despite the increasing deployment of DMS, there is no dedicated ethical framework guiding their development and integration. This absence creates legal ambiguity, potential regulatory pushback, and a lack of public confidence - threatening to delay widespread implementation and diminish potential benefits \citep{garikapati2024autonomous, vellinga2021automated}. As DMS become more central to AV ecosystems, they introduce complex questions around liability, acceptable intervention thresholds, and individual rights in the context of continuous monitoring. Legal systems have struggled to keep pace with these rapidly evolving technologies. 

\begin{figure}
    \centering
    \includegraphics[width=1\linewidth]{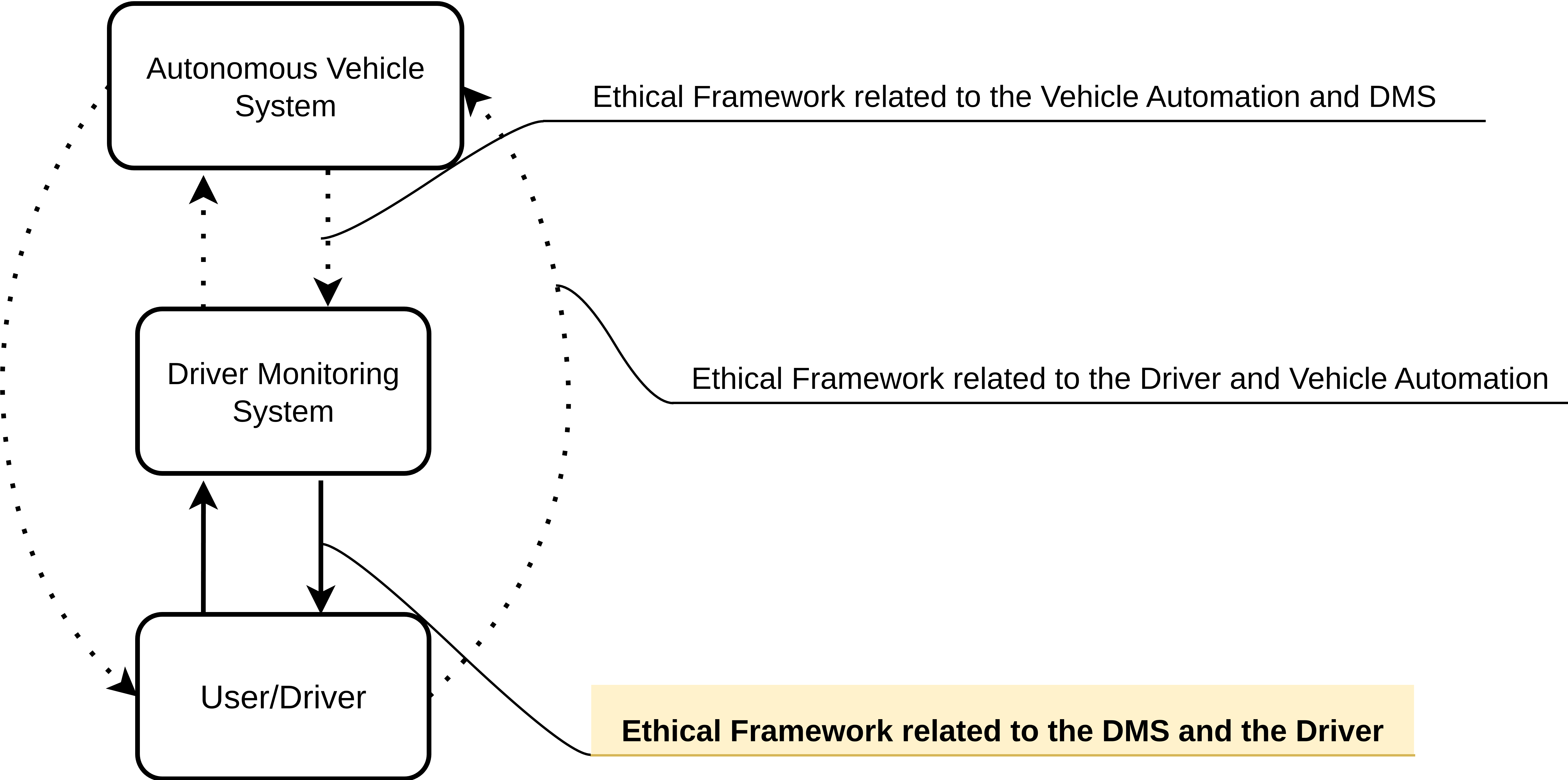}
    \caption{Categorisation of Ethical Framework based on different interfaces, e.g. AVs and DMS, DMS and Driver, Driver and AVs}
    \label{fig:figure1}
\end{figure}

Although various studies have attempted to address the ethical implications of AVs and monitoring technologies \citep{cahill2020advancing, cahill2020ethical}, these efforts remain fragmented. The lack of a cohesive, interface-specific framework has limited the effectiveness of current policy and system design recommendations. This paper addresses that gap by identifying three key interfaces, as shown in figure\ref{fig:figure1}, and organizing ethical challenges at the relevant interface levels: (i) between AVs and DMS, (ii) between DMS and the human driver, and (iii) between AVs and the human driver. Table \ref{tab:table1} discusses key ethical questions related to these interfaces.

\begin{table}[htbp]
\centering
\small % reduce font size for the whole table
\setlength{\tabcolsep}{3pt} % less horizontal padding
\caption{Ethical challenges and key questions related to AVs across AV--DMS--driver interfaces}
\label{tab:table1}
\begin{tabular}{|p{3.3cm}|p{4.1cm}|p{6.1cm}|}
\hline
\textbf{Interface} & \textbf{Challenge} & \textbf{Key Ethical Questions} \\
\hline
\multirow{4}{*}{AV and DMS}
  & Accuracy and reliability
  & Can we trust the DMS to make safe, real-time decisions based on sensor input? \\
  & Decision-making autonomy
  & How should decision-making responsibility be shared between the AV system and the DMS? \\
  & Algorithmic transparency
  & Are the algorithms powering the AV and DMS transparent, explainable, and auditable? \\
  & Ethical design of intervention protocols
  & When and how should the AV act on data collected by the DMS (e.g., slowing down, pulling over)? \\
\hline
\multirow{3}{*}{\shortstack{DMS \\ and Human Driver}}
  & Privacy and data protection
  & Who owns the data collected by the DMS, and how is it stored, processed, and shared with third parties? \\
  & Consent and informed usage
  & Are drivers fully aware of the extent of in-cabin monitoring, and have they provided informed, revocable consent? \\
  & Psychological burden of surveillance
  & How does constant monitoring affect drivers’ autonomy, perceived fairness, and long-term trust in the system? \\
\hline
\multirow{3}{*}{\shortstack{AV \\ and Human Driver}}
  & Responsibility and accountability
  & Who is liable in case of a crash — the driver, the AV system, or the DMS provider/integrator? \\
  & Role ambiguity
  & Does the driver clearly understand their residual role and when they are expected to take over? \\
  & Takeover request ethics
  & Are drivers given adequate time, modality, and information to safely regain control of the vehicle? \\
\hline
\end{tabular}
\end{table}

This layered categorization offers a more effective understanding of the ethical landscape and reveals the interplay between responsibility, autonomy, and surveillance in intelligent mobility systems (see figure ~\ref{fig:figure1}). To the best of our knowledge, this interface-based distinction has not been clearly articulated in existing literature, where DMS ethics are often subsumed under broader discussions of AV ethics. Our approach echoes the European Commission’s ethical guidelines for AVs, which organize ethical issues into domains such as road safety, data ethics, and responsibility - seeking to disentangle overlapping ethical concerns for more targeted governance. 

\begin{figure}
    \centering
    \includegraphics[width=0.8\linewidth]{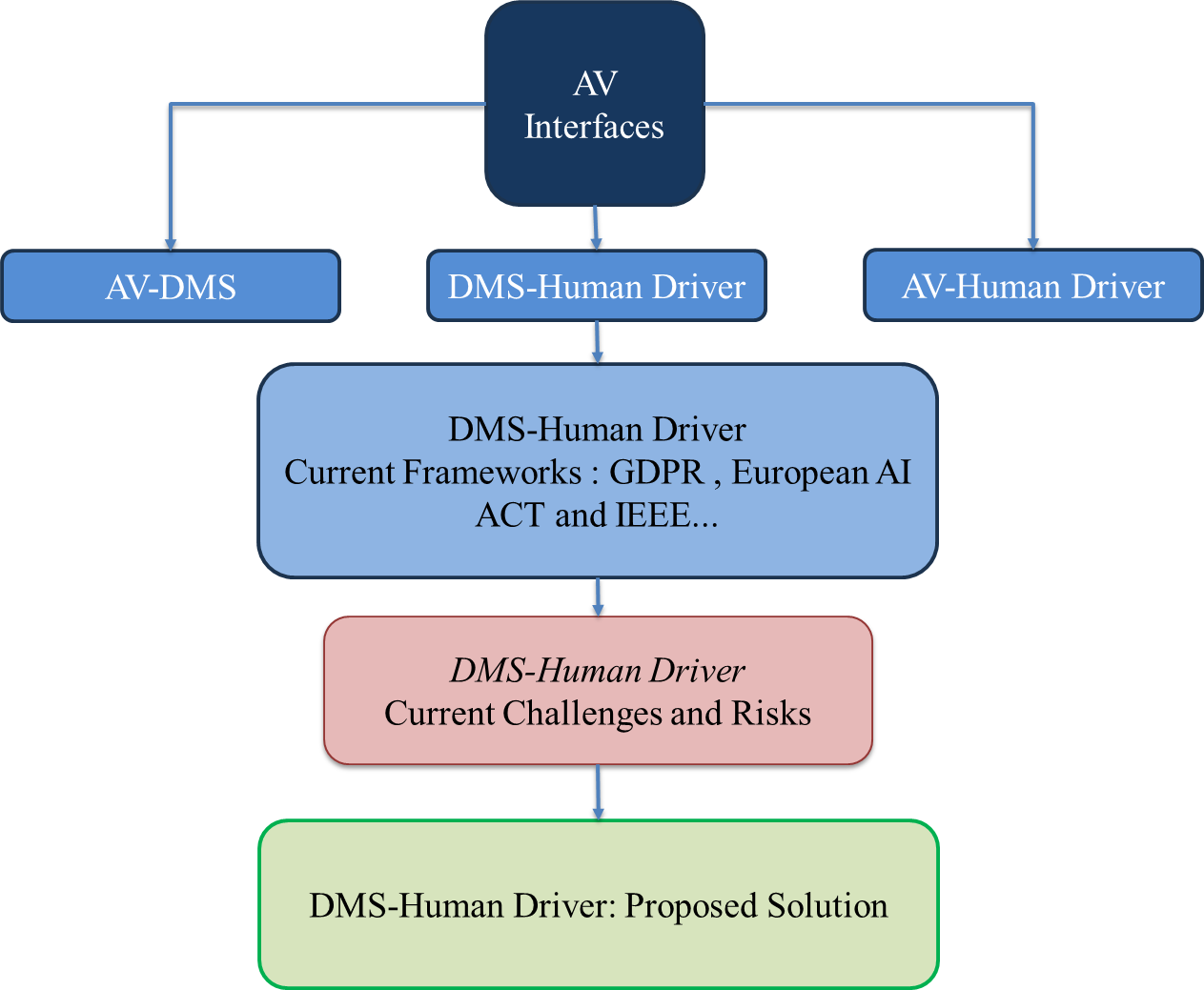}
    \caption{Scope of this paper}
    \label{fig:figure2}
\end{figure}

\subsection{Scope of this Paper}

This paper focuses exclusively on the ethical challenges at the interface between DMS and the human driver as described in figure \ref{fig:figure2}. While AV ethics are widely discussed, the specific ethical, legal, and psychological issues related to in-cabin monitoring have received limited dedicated analysis. We address this gap by examining the DMS-driver interaction as a standalone domain deserving focused attention.

There are two primary reasons for narrowing our scope to this interface. First, unlike many speculative AV scenarios, DMS technologies are already in widespread deployment and are becoming mandatory in regions like the EU \citep{euro2022euro}. This makes their ethical impact immediate, practical, and policy relevant. Second, DMS collect and process highly sensitive biometric and behavioural data such as gaze, facial expressions, and cognitive state raising serious concerns around privacy, consent, data protection, and user autonomy.

Focusing on this interface also avoids conflating DMS ethics with broader AV dilemmas and allows for a deeper analysis of real-world ethical risks. It aligns with findings from \citep{coyne2024understanding}, which highlight that drivers often perceive DMS and automated driving systems as distinct and uncoordinated. Our analysis aims to clarify the ethical foundations of in-cabin monitoring and provide guidance for system designers, regulators, and policymakers. Furthermore, by being automation level agnostic (as per SAE definitions) \citep{sae2021sae}, our framework remains relevant across all stages of vehicle autonomy from advanced driver assistance systems to fully autonomous vehicles.

\section{Related Regulations/Framework (IEEE, EU AI Act, \& GDPR) }
As DMS continue to evolve, it relies increasingly on the collection and processing of personal and biometric data. This dependence raises complex ethical and legal challenges, especially around data privacy, transparency, and accountability. To address these concerns, several regulatory and ethical frameworks have been introduced to guide responsible AI development in DMS contexts. Key among these are the General Data Protection Regulation (GDPR), the EU Artificial Intelligence Act (EU AI Act), and the IEEE’s Ethically Aligned Design (EAD) \citep{garikapati2024autonomous, cunningham2015autonomous, sae2021sae, coyne2024understanding, jatavallabha2024tesla, visconti2025innovative, zuboff2015big}.  

\subsection{General Data Protection Regulation (GDPR) }
The GDPR, implemented by the European Union in 2018, marked a foundational shift in global data governance practices \citep{garikapati2024autonomous}. It emphasizes key principles such as explicit user consent, data minimization, and purpose limitation - principles especially pertinent to DMS, which process sensitive biometric data like facial expressions, cardiovascular indicators, and eye movements. The classification of such data as “special categories” under GDPR means it can only be processed with strict safeguards, including legitimate purpose and informed consent (Article 5, Section 1.b; Article 6) \citep{garikapati2024autonomous}. 

The regulation has proven to be a strong model, influencing legislation in other regions such as Canada's Personal Information Protection and Electronic Documents Act (PIPEDA) and Australia’s Intelligent Transport Systems (ITS) data protection framework\citep{sae2021sae},\citep{jatavallabha2024tesla}. Its impact is further underscored by enforcement actions like the €50 million fine issued to Google in 2019 for non-compliance \citep{cunningham2015autonomous}, highlighting the EU’s readiness to uphold the regulation with substantial penalties. 

Despite its protective intent, GDPR’s rigorous requirements have been viewed by some as obstacles to rapid innovation, particularly in data-intensive fields such as DMS development. The complexity of compliance and the prohibition against third-party sharing - reinforced by the General Safety Regulation (GSR) enforced from July 2022 - can delay system integration and limit data availability for machine learning and safety improvement purposes \citep{zuboff2015big}. Nonetheless, GDPR remains a cornerstone in safeguarding the privacy rights of EU residents in the deployment of intelligent mobility systems. 

\subsection{EU Artificial Intelligence Act (EU AI Act) }
The EU Artificial Intelligence Act, adopted in 2024, represents the first major legislative framework focused specifically on the regulation of AI technologies \citep{act2024eu}. It introduces a risk-based categorization of AI systems - ranging from minimal to high risk - tailoring legal obligations accordingly. The Act mandates that high risk systems undergo rigorous risk assessments, maintain high standards of transparency and robustness, and include mechanisms for human oversight \citep{hayley2021driver}. This structured approach allows for more targeted regulation compared to broader instruments like the GDPR. It also operationalizes ethical principles such as fairness, accountability, and human-centricity - principles central to the responsible deployment of AI in mobility contexts \citep{coyne2024understanding}. 

In practical terms, compliance with the EU AI Act necessitates an integrated governance approach that spans the entire lifecycle of the DMS. This includes systematic risk management, human-in-the-loop control mechanisms, and detailed technical documentation that can be audited by regulators or conformity assessment bodies. Importantly, the AI Act also emphasizes traceability and accountability - ensuring that automated decisions affecting human safety can be explained and contested.

For academic and industrial research groups, this framework encourages the adoption of \textit{ethics-by-design} and \textit{compliance-by-design} methodologies early in system development. Aligning model development, data handling, and testing protocols with the AI Act’s transparency and documentation requirements can streamline later conformity assessments and foster public trust in AI-enabled mobility solutions. While these processes introduce additional administrative and technical burdens, they also create opportunities to establish best practices for safe and ethical AI deployment in semi-automated and fully autonomous vehicle ecosystems.

\subsection{IEEE Ethically Aligned Design (EAD)}
The IEEE’s Ethically Aligned Design (EAD) initiative offers a complementary, non-legislative framework that emphasizes ethical considerations throughout the AI development lifecycle \citep{visconti2025innovative}. Unlike legally binding regulations, IEEE EAD serves as a set of aspirational guidelines promoting value-sensitive design, social impact assessments, and the inclusion of diverse stakeholders in the design process. This framework is particularly relevant to DMS technologies, which have implications beyond technical functionality - such as the psychological and behavioural effects of continuous driver surveillance \citep{visconti2025innovative}. IEEE EAD encourages developers to evaluate these less tangible consequences and to embed ethical reflection into the earliest stages of design. It thereby extends ethical governance from a compliance-based approach to one that is anticipatory and human-centred. 

Nevertheless, the voluntary nature of the IEEE EAD framework means that its adoption is not guaranteed across the industry. Without formal enforcement mechanisms, the responsibility for ethical integrity falls primarily on individual developers and organizations. Moreover, translating high-level ethical values - such as dignity or well-being - into measurable engineering criteria can be complex, especially in competitive commercial settings. Still, the IEEE EAD initiative plays a vital role in broadening the conversation around ethics in AI beyond mere legality. 

\section{DMS-Human Driver Interface: Ethical Challenges and Risks}
\label{sec:E-challenges}

DMS are in-cabin sensing technologies that assess driver state and behaviour in real-time to enhance safety in partially and conditionally (Level 2-5 as per SAE) automated vehicles \citep{sae2021sae}. Leveraging visual, auditory, and physiological sensors, DMS detect signs of drowsiness, distraction, cognitive overload, and medical emergencies. With growing regulatory mandates (e.g., EU GSR), these systems are rapidly evolving, integrating AI and multi-modal sensing to provide robust insights into driver and occupant status. 

By continuously evaluating driver readiness and enabling timely interventions, DMS help mitigate human error - the leading cause of road accidents - enhance overall driving safety, and foster user confidence in vehicle technologies.  

\begin{table*}[htbp]
\centering
\scriptsize
\renewcommand{\arraystretch}{1.2}
\begin{tabularx}{\textwidth}{
>{\raggedright\arraybackslash}p{2.8cm}
X X X X X}
\toprule
\textbf{Feature} & 
\textbf{RGB-Based} & 
\textbf{Radar Sensing} & 
\textbf{Infrared (IR)/Thermal} & 
\textbf{Event-Based Cameras} & 
\textbf{Audio-Based} \\
\midrule

\textbf{Tech Type} &
RGB cameras, gaze estimation &
60 GHz FMCW radar (Infineon, Vayyar, TI), MIMO radar arrays &
IR cameras, LWIR/MWIR, thermopile arrays, passive infrared imaging &
Asynchronous pixel sensors; capture per-pixel brightness changes &
MEMS microphones, speech emotion AI \\
\midrule

\textbf{Main Applications} &
Drowsiness, distraction, hands-on-wheel detection, driver ID, behavior profiling &
Child presence detection, respiration \& heartbeat, motion/intrusion detection &
Drowsiness detection via facial/eye temperature, occupant presence, fatigue, emotion estimation &
Eye blink detection, microsaccades, sudden head movements &
Occupant classification via voice, voice-based alertness/emotion detection \\
\midrule

\textbf{Hardware Examples} &
Sony IMX, Luxonis OAK-D, Omnivision OV2311 &
Infineon BGT60TR13C, Vayyar 4D MIMO sensor &
FLIR Lepton, FLIR Boson, Workswell WIRIS, Optris PI series &
Prophesee EVK, iniVation DAVIS &
ReSpeaker Mic Arrays, Knowles MEMS \\
\midrule

\textbf{Strengths} &
Effective in various lighting; scalable from L1–L4; embedded or centralized processing &
Contactless; works in darkness; can monitor through soft materials &
Passive and privacy-preserving; works in total darkness; captures physiological heat &
Ultra-low latency; high temporal resolution; robust under extreme lighting &
Voice-driven analytics; cost-effective for cabin classification \\
\midrule

\textbf{Limitations} &
Obstructed view (e.g., sunglasses); requires facial visibility &
Lower spatial resolution; struggles with stationary targets &
Lower resolution/contrast; sensitive to thermal noise; higher cost &
Newer tech; requires specialized processing; not yet widely deployed &
Noise-sensitive; needs advanced filtering \\
\bottomrule

\end{tabularx}
\caption{Comparison of In-Cabin Sensing Modalities}
\label{tab:sensing_comparison}
\end{table*}

% Before we dig into the ethical aspects of the DMS-Human-driver, we will first explore the sensing modalities currently involved in DMS-incabin data acqusition. 
\subsection{Sensing Modalities in Driver Monitoring Systems}

DMS use a range of sensing modalities to assess driver state, behaviour, and wellbeing. Each sensor type offers specific advantages for detecting fatigue, distraction, or cognitive overload, but also introduces unique technical and ethical challenges. This section categorizes and discusses the key modalities commonly used in modern DMS deployments. Table \ref{tab:sensing_comparison} further discusses and compares different sensor technologies.

\subsubsection{Frame-Based Vision: RGB, IR, and Thermal Cameras}

Frame-based vision systems, particularly RGB and infrared (IR) cameras, are the most established and widely deployed sensing modality in production vehicles. These systems monitor facial landmarks to estimate gaze direction, eye closure, head pose, and other indicators of drowsiness or distraction. IR illumination enables functionality in low-light conditions, ensuring compliance with standards such as Euro NCAP requirements for detecting secondary tasks like phone use or eating \citep{magdalena2018sparseppg, eventDMS}.

Thermal imaging, or long-wave infrared (LWIR), detects heat signatures from the driver’s face and body to assess physiological cues such as fatigue, stress, and breathing. It is also used for child presence detection and general occupant monitoring. Unlike RGB and IR, thermal sensors are less affected by ambient light, occlusions, or individual appearance variations, and they offer some privacy advantages due to the abstract nature of the data \citep{thermalreview}.

Despite their effectiveness, vision-based systems face challenges including occlusions (e.g., sunglasses), varying lighting conditions, inter-user variability, and privacy risks from detailed facial imagery. Continuous recording without explicit consent raises regulatory concerns, especially under data protection laws like the GDPR. Thermal imaging adds concerns around health-related inferences and non-consensual physiological monitoring \citep{thermalreview}.

\subsubsection{Event-Based Vision (Asynchronous Cameras)}

Event cameras offer a novel approach to visual sensing by asynchronously capturing only pixel-level brightness changes, rather than full frames \citep{event-review}. This results in high temporal resolution and low latency, making them well-suited for detecting subtle or rapid driver movements such as microsaccades, blinks, or rapid head turns that may not be captured by conventional cameras \citep{yawn-event}.

Due to their sparse data output and low visual fidelity, event cameras may offer enhanced privacy compared to traditional imaging. However, their novelty introduces challenges related to data interpretation, algorithmic fairness, standardization, and integration within existing legal frameworks. Their use in DMS remains largely in the research and prototyping phase \citep{dd-eventcamera, eventDMS, yawn-event, event-review}.

\subsubsection{Radar-Based Sensing (60 GHz)}

60 GHz radar sensors detect fine physiological movements including heartbeats, and respiration, etc. In DMS, they are primarily used for child presence detection, occupancy monitoring, and emerging applications in driver wellness and fatigue detection. Radar offers advantages such as robustness to lighting conditions, ability to function through soft obstructions (e.g., clothing, blankets), and preservation of visual privacy \citep{jung2021cnn}.

However, radar systems are relatively costly, require advanced signal processing, and may generate false positives due to environmental noise. These sensors are increasingly being integrated into premium vehicles by manufacturers such as Tesla and Toyota, and are expected to play a larger role in health-centric automotive applications. With wider adoption and mass production, their costs have been decreasing steadily, making them more accessible for mid-range vehicles as well \citep{jung2021cnn}.

\subsubsection{Audio-Based Sensing (Microphones)}

Acoustic sensing complements vision-based DMS by capturing both speech and non-speech auditory cues from the driver and cabin environment. Beyond conventional voice commands, these systems analyze para-linguistic signals - such as yawning, coughing, humming, or groaning - to infer physiological or cognitive states. Variations in speech patterns, vocal tone, and prosodic features can further reveal indicators of cognitive load, stress, fatigue, or distraction, thereby providing an additional non-visual modality for driver state assessment \citep{jain2023adaptation, jain}. They can also support voice-based control interfaces \citep{speech-gianGonzales}.

Audio sensors function well in low-light or visually obstructed conditions and offer a cost-effective complement to visual systems. However, they raise substantial privacy and legal concerns, especially around the recording of speech and emotion recognition. In some jurisdictions, processing emotional data from voice may face strict legal scrutiny. These systems are commonly used as supplementary modalities in commercial platforms like Nauto and Samsara.

\subsection{Ethical Challenges and Risks related to DMS-Driver interface }
While ethical debates around autonomous vehicles often focus on fully driverless systems and high-stakes moral dilemmas, most AVs currently operate at intermediate levels - specifically SAE Levels 3 and 4 - where human supervision remains essential \citep{sae2021sae}. In these settings, the vehicle can manage driving under certain conditions but still requires the driver to take over when prompted. This shared control raises its own ethical concerns, particularly around driver readiness and accountability during transitions. DMS have emerged as a crucial tool to address these issues by assessing driver attention and alertness in real time. However, while DMS can enhance safety, they also raise significant ethical questions related to privacy, autonomy, and psychological impact - making them both a solution and a new ethical frontier within the AV landscape. 

A recent qualitative study by \citep{coyne2024understanding} found that drivers tended to view DMS more positively than Automated Driving Systems (ADS), however, they remained wary of several key issues, particularly in terms of data privacy, surveillance, system reliability, and the possibility of personal information being shared or exploited. There was also concern that such monitoring could could result in a less enjoyable driving experience. Similarly, in another study \citep{presta2022would}, the authors explored driver acceptance of Brain Computer Interface (BCI)-based versus non-invasive DMS. The research indicated that drivers exhibited significant privacy concerns regarding BCI-based DMS, primarily due to the invasive nature of neural data collection. In contrast, non-invasive systems are generally more acceptable to drivers, as they are perceived to be less intrusive and more respectful of personal privacy.  These findings point to a broader need for ethical safeguards and transparent communication in the development of in-vehicle monitoring technologies.   

Several studies have identified a range of ethical concerns has emerged \citep{coyne2024understanding, bhoopalam2023long, yawn-event, dd-eventcamera, lee2004trust}. These issues span privacy, consent, algorithmic bias, transparency, data governance, and psychological well-being. Table \ref{tab:ethical_considerations} provides some ethical questions related to each of these mentioned categories.

\begin{table*}[htbp]
\centering
\scriptsize
\renewcommand{\arraystretch}{1.25}
\begin{tabularx}{\textwidth}{
>{\raggedright\arraybackslash}p{3.3cm}
>{\raggedright\arraybackslash}p{4.5cm}
X}
\toprule
\textbf{Domain} & \textbf{Challenge} & \textbf{Key Ethical Questions} \\
\midrule

\textbf{Privacy and Surveillance} &
Continuous monitoring may be perceived as intrusive. &

• Is the driver aware of the extent and purpose of monitoring? \newline
• Can the driver opt out of certain types of data collection? \newline
• How is sensitive data anonymized or encrypted? \\
\midrule

\textbf{Consent and Autonomy} &
Drivers may lack control or meaningful consent over DMS functionality. &
• Is the driver’s consent actively sought before system activation? \newline
• Can features be disabled without compromising safety? \\
\midrule

\textbf{Data Ownership and Sharing} &
Ownership and access policies are often unclear. &
• Who owns the collected data — the driver, carmaker, or third party? \newline
• Is data shared with insurers, law enforcement, or advertisers? \newline
• Are policies in place that restrict data misuse? \\
\midrule

\textbf{Bias and Fairness} &
Algorithms may not work equally well across demographics. &
• Is the system accurate for all ages, genders, ethnicities, and abilities? \newline
• Are there mechanisms to detect and correct algorithmic bias? \\
\midrule

\textbf{Transparency and Explainability} &
Drivers may not understand how the system makes decisions. &
• Can drivers understand why alerts or interventions occur? \newline
• Are logs and decisions auditable by humans? \\
\midrule

\textbf{Security and Misuse of Data} &
Risk of hacking, leaks, or misuse beyond safety. &
• What cybersecurity measures are in place? \newline
• What are the consequences of a data breach? \\
\midrule

\textbf{Health-Related Detection and Intervention} &
Raises ethical dilemmas around health data and system actions. &
• Can the system distinguish between critical and non-critical health issues? \newline
• What actions should be taken when health anomalies are detected? \newline
• What are the ethical or legal duties if the system fails during a health emergency? \\
\midrule

\textbf{Psychological Impact} &
Constant monitoring may affect driver well-being. &
• How does continuous monitoring affect stress or behavior? \newline
• Can system sensitivity be personalized to reduce discomfort? \\
\bottomrule

\end{tabularx}
\caption{Ethical Considerations in Driver Monitoring Systems (DMS)}
\label{tab:ethical_considerations}
\end{table*}

\subsubsection{Privacy and Surveillance }
DMS often collect highly sensitive behavioural and physiological data - such as eye gaze, facial expressions, and biometric signals like heart rate or blink patterns - to detect driver physical and cognitive states \citep{khan2019comprehensive}. While this enhances safety, it introduces a significant risk of infringing on users’ privacy. One major concern is the lack of transparency: drivers are often unaware of what data is being collected, how it is stored, and who can access it. The omnipresence of in-cabin cameras and sensors may lead to a "panoptic effect", where individuals feel constantly observed, potentially altering their behaviour \citep{johnson2014you, li2024panoptic}. This discomfort is amplified by broader concerns around surveillance, especially when coupled with the fear that collected data could be misused or shared with third parties. 

Public scepticism is not completely unfounded - several U.S. states, including California, New York, and Massachusetts, have enacted bans or restrictions on facial recognition technologies, citing privacy and misuse concerns \citep{almeida2022ethics}. These regulatory moves also reflect growing societal discomfort with technologies that track and analyse personal features without explicit consent. 

\subsubsection{Consent and Autonomy }
A foundational ethical principle in technology use is informed consent. However, in many DMS implementations, consent is often implicit or buried within lengthy end-user agreements, limiting drivers’ awareness and control over what data is collected, how it is used, and with whom it is shared \citep{sucharski2016privacy, Barry2025-bbb}. This undermines the principle of consent as defined by regulations like the EU GDPR, which requires consent to be freely given, specific, informed, and unambiguous \citep{regulation2018general}.

In practice, drivers may not realize that their biometric and behavioural data - such as gaze, facial expressions, or physiological signals - are being continuously monitored. Consent is frequently bundled into general terms and conditions, reducing transparency and diminishing user agency. Furthermore, disabling DMS features may impair critical safety functions or conflict with regulatory mandates, creating a trade-off between compliance and autonomy.

Continuous in-cabin monitoring can also affect the driver's psychological state. Studies report that persistent surveillance can lead to discomfort, self-censorship, or over-reliance on automation, potentially reducing situational awareness and intervention readiness during takeovers \citep{lee2004trust, coughlin2011monitoring, coyne2024understanding, bhoopalam2023long, yawn-event, dd-eventcamera, coughlin2011monitoring}.

These concerns are echoed in qualitative research, where drivers express scepticism about DMS, fearing their data could be misused - such as being shared with insurers or used to assign blame in accidents \citep{vellinga2021automated} \citep{coyne2024understanding}. Many feel they lack control over what is monitored and how it is interpreted.

To uphold ethical standards, DMS developers and regulators must implement transparent, user-friendly consent mechanisms and provide meaningful control options - without compromising safety-critical functionality \citep{hayley2021driver}.

\subsubsection{Data Ownership and Sharing }
One of the most pressing and unresolved ethical concerns surrounding DMS is the question of data ownership. DMS collect a wide range of behavioural and biometric data - from eye gaze to facial expression variability and cognitive workload indicators. However, there remains considerable ambiguity regarding who rightfully owns this data - the driver, the vehicle manufacturer, the DMS provider, or possibly even third-party entities \citep{costantini2020autonomous}. 

This lack of clear data governance creates vulnerabilities. Without firm regulatory boundaries or contractual transparency, data may be accessed or shared with external stakeholders such as insurance companies, law enforcement, advertisers, or even data brokers, often without the driver’s explicit knowledge or consent \citep{gaeta2019data}. 
Studies have shown that most drivers consider it unacceptable for their in-vehicle data - particularly biometric or behavioural metrics - to be shared with third parties \citep{josten2017privacy} This perception reflects a strong public expectation of informational autonomy and reinforces the need for consent-centric data practices.

Such unauthorized or opaque data exchanges raise substantial concerns about informational self-determination, particularly in jurisdictions with weak or uneven enforcement of data protection regulations. For instance, drivers may find themselves at a disadvantage if DMS data is used to assign liability in crash investigations, calculate insurance premiums, or generate behavioural profiles that could affect their creditworthiness or employment prospects. Such uses go far beyond the original purpose of ensuring safety and erode public trust in these technologies \citep{vellinga2021automated}. 

From a legal standpoint, frameworks such as the GDPR attempt to clarify rights over personal data, including the right to access, rectify, and erase it \citep{regulation2018general}. However, applying these rights in the context of AVs remains a technical and regulatory challenge, particularly when data is continuously streamed and processed by cloud-based services with multiple intermediaries \citep{shelby2023sociotechnical}. 

\subsubsection{Bias and Fairness }
Bias and fairness are critical ethical concerns in the deployment of a DMS. These technologies, particularly those relying on computer vision and facial recognition, may not perform equally across diverse demographic groups, skin types, and ethnicities, etc. \citep{almeida2022ethics}. Studies have revealed that facial analysis algorithms often underperform for individuals with darker skin tones, non-Western facial features, or those wearing cultural or religious head coverings, leading to inaccurate assessments of attention or fatigue levels \citep{shelby2023sociotechnical, jambholkar2024ethical}. Such biases can have real-world consequences: drivers may be unfairly flagged as inattentive or impaired, leading to increased stress, wrongful interventions, or even penalization \citep{jambholkar2024ethical}. These systemic inaccuracies not only undermine user trust but also amplify social inequities in how safety technologies are applied. A notable example is Twitter’s 2021 controversy, in which the platform’s image-cropping algorithm was found to exhibit racial and gender bias - favouring lighter-skinned and male faces. The incident triggered widespread backlash and prompted the company to publicly commit to implementing six core principles of AI ethics, including fairness and inclusivity \citep{Clark2025-mmm}. 

Similar issues have been reported in other AI contexts, such as hiring systems and healthcare algorithms, which have been criticized for perpetuating historical biases in training data \citep{hofeditz2022applying}. These examples underscore the importance of auditing datasets for demographic representativeness and developing fairness-aware learning algorithms.  

\subsubsection{Transparency and Explainability }
A significant ethical concern surrounding DMS is their lack of transparency and explainability. Many AI-driven DMS solutions operate as “black boxes,” offering minimal insight into the decision-making processes behind alerts or interventions \citep{Sultana2024-ppp, ryan2019semiautonomous, koesdwiady2016recent}. This opacity can result in user confusion, diminished trust, and frustration - particularly when drivers are unable to understand why they have been flagged or corrected \citep{Blake2024-qqq} \citep{cunneen2019autonomous}. Explainability is critical for ensuring accountability and user acceptance, especially in contexts where safety and behavioural assessments are involved \citep{Blake2024-qqq}.  

\subsubsection{Security and Misuse of Data }
Given the highly sensitive nature of data collected by DMS - including biometric, behavioural, and physiological indicators - it is critical that robust cybersecurity measures are in place. A compromised DMS poses serious risks, such as real-time surveillance, unauthorized tracking, blackmail, and unconsented data sharing, all of which could harm individuals both professionally and personally \citep{ryan2019semiautonomous} \citep{koesdwiady2016recent}. 

Ethical DMS design must therefore prioritize data encryption, secure storage, and well-defined data breach response protocols \citep{sargiotis2024data} \citep{nasir2024ethical}. Equally important are clear limitations on data retention and strong access controls that prevent misuse by internal or external actors \citep{regulation2018general}. While legal instruments such as the GDPR offer guidance on consent, data subject rights, and data controller responsibilities, gaps remain in how these are applied to Autonomous and ACS \citep{thermalreview}. Studies highlighted the importance of Data Protection Impact Assessments (DPIA) and privacy-preserving processing principles across different data ecosystems \citep{bu2020controller} \citep{mulder2021exploring}. However, their scope remains broad, not accounting for the unique context of in-vehicle systems.

\subsubsection{Health-Related Detection and Intervention }

Advanced DMS are increasingly capable of detecting early signs of health conditions such as drowsiness, micro-sleeps, and even acute medical emergencies like cardiac arrest or stroke. While such features can significantly enhance road safety and even save lives, they also introduce ethical and legal dilemmas regarding system autonomy and the appropriate course of action \citep{hayashi2021toward}\citep{cahill2020advancing}. 

Key Ethical Questions include: Should the DMS intervene autonomously, for example by stopping the vehicle or alerting emergency services? And, what are the implications if it either fails to intervene in a genuine emergency or mistakenly triggers a response in a non-critical event? These scenarios raise concerns around false positives, liability, and the scope of machine autonomy in medical contexts. 

\subsubsection{Psychological Impact }

Continuous driver monitoring may lead to psychological discomfort, increased stress, or anxiety, especially if users perceive the system as punitive or intrusive \citep{bhoopalam2023long}. This is particularly concerning for professional drivers, who often operate under strict regulations and time pressure, making them more susceptible to increased mental workload \citep{greenfield2016truck}. Studies suggest that drivers may experience distrust in the technology, reduced autonomy, and self-consciousness due to the feeling of being constantly watched.

\section{Ethical Principles Guiding Framework Design}
\label{sec:principles}
Before presenting the proposed framework, it is important to first examine the core ethical principles and guidelines that will inform its development. This section outlines the normative foundations relevant to DMS, including key considerations such as data subject rights, privacy, ownership, consent, and accountability. These principles are not tied to a specific technical implementation but serve as essential reference points for any ethical evaluation of DMS technologies. Establishing these values upfront ensures that the eventual framework is grounded in a transparent, stakeholder-aware, and rights-respecting foundation.

The framework is:
\begin{itemize}
    \item Principle-based: Anchored in widely recognized ethical values such as autonomy, fairness, and accountability. 
    \item Modular and adaptable: Capable of being tailored to different deployment contexts, system architectures, and legal environments. 
    \item Open to revision: Designed to evolve through stakeholder feedback, future iterations, and implementation experience. 
\end{itemize}

\begin{figure*}[htbp]
    \centering
    \includegraphics[width=1.0\textwidth]{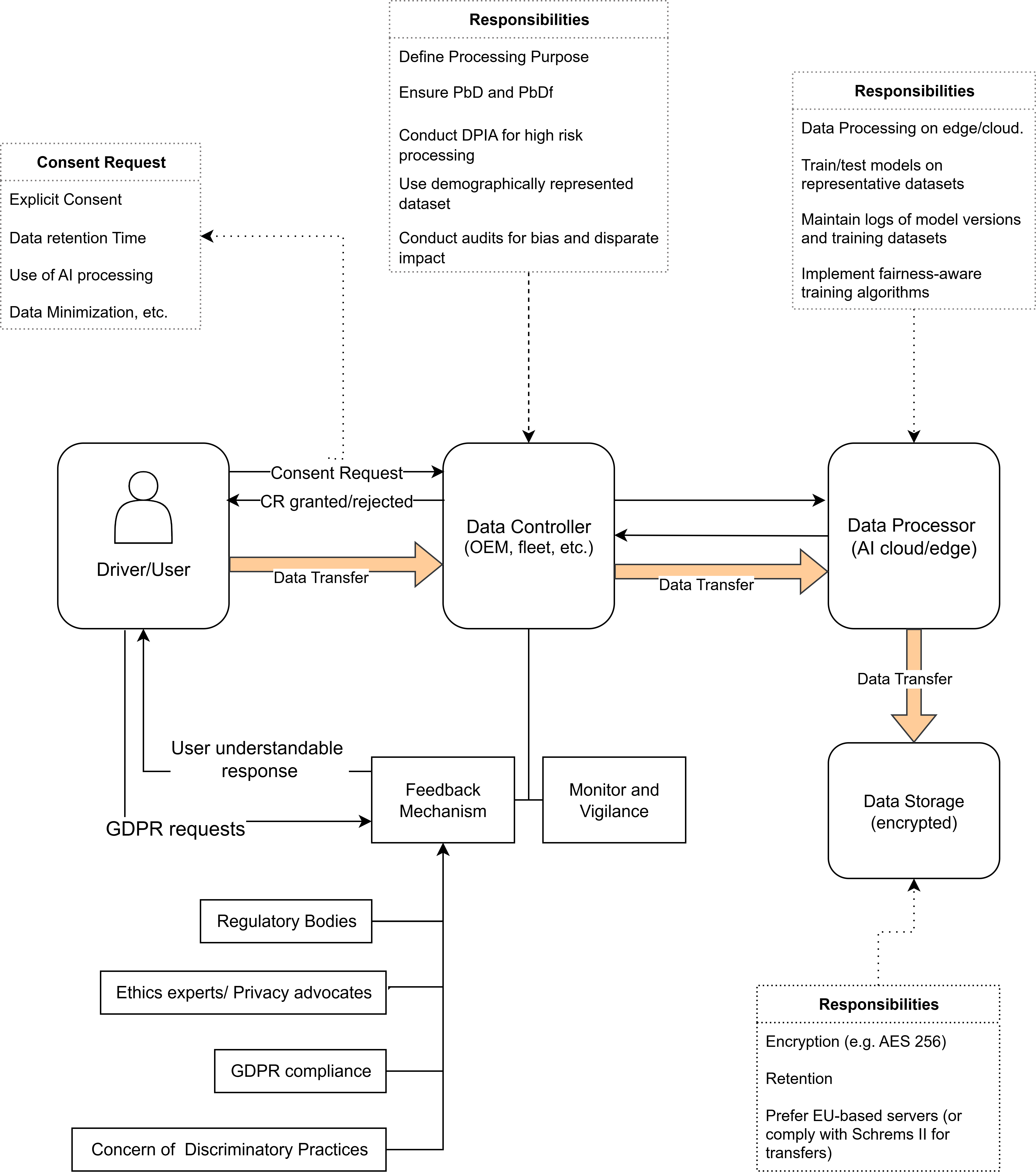}
    \caption{Proposed ethical framework for DMS showing GDPR-compliant data stewardship and oversight among key stakeholders.}
    \label{fig:Figure3}
\end{figure*}

Drawing on the conceptual framework for ethical AI development in IT systems draws upon a range of multidisciplinary perspectives, synthesizing insights from ethics, computer science, law, philosophy, and other relevant fields. It aims to address the complex ethical, legal, and social challenges posed by AI technologies \citep{Olorunfemi2024-aaaa}. In light of the transparency, explainability, and data protection imperatives discussed above, we propose a structured ethical framework to guide the responsible development and deployment of DMS in AVs. This framework not only complements legal compliance (e.g., with GDPR and EU AI act) but also addresses broader concerns of trust, fairness, and human dignity in AI-enabled mobility systems. This ethical framework illustrated in figure\ref{fig:Figure3} and \ref{fig:Figure4} provides a comprehensive approach to managing key ethical concerns associated with DMS. By integrating legal mandates such as the GDPR and aligning with principles from the EU AI Act, the framework addresses multiple domains of ethical risk \citep{regulation2018general}. 

Specifically, the framework incorporates the GDPR’s Data Processing Principles, including: 

\begin{itemize}
    \item \textbf{Lawfulness, Fairness, and Transparency} (Article 5(1)(a), Article 6)
    \item \textbf{Purpose Limitation} (Article 5(1)(b))
    \item \textbf{Data Minimization and Storage Limitation} (Article 5(1)(c-e))
    \item \textbf{Accuracy of Data} (Article 5(1)(d))
    \item \textbf{Integrity and Confidentiality} via Privacy by Design and Privacy by Default (Article 25(1-2))
    \item \textbf{Accountability Mechanisms} to demonstrate compliance (Article 5(2))
\end{itemize}

These additions ensure that DMS respect individual autonomy while maintaining system efficacy and public trust. By embedding ethical considerations into both the technical architecture and governance structure of DMS, this framework serves as a comprehensive guide for the responsible integration of AI in mobility systems.

% \section{Framework Response to Ide Ethical Framework for Driver Monitoring Systems (DMS)}
% % This section presents a provisional ethical framework for DMS/ICS, grounded in the ethical challenges identified in Section 3.2 and informed by relevant regulatory and normative frameworks (Section 2.2). The aim of this framework is not to prescribe rigid technical or policy solutions but rather to offer a structured basis for discussion and co-creation among stakeholders, including industry partners, researchers, regulators, and end users. It is intended to be: 

% This section introduces an ethical framework for DMS, informed by the ethical challenges identified in Section \ref{sec:E-challenges} and aligned with regulatory principles outlined in Section \ref{sec:principles}. 

\section{Proposed Framework}

This section proposes ethical framework based on the principles identified in the section \ref{sec:principles}. The aim of this framework is not to prescribe rigid technical or policy solutions but rather to offer a structured basis for discussion and co-creation among stakeholders, including industry partners, researchers, regulators, and end users.

Figure \ref{fig:Figure3} illustrates the proposed ethical framework for Driver Monitoring Systems (DMS), which embeds a data stewardship model designed to ensure transparency, accountability, and GDPR compliance in automated vehicle ecosystems. The framework traces the entire lifecycle of driver data - from collection and consent to processing, storage, and oversight - emphasizing user rights and ethical responsibility at every stage.

At the entry point of the framework, the Driver or User acts as the data subject. Any in-cabin sensing or biometric data collection begins only after a clear consent request (CR) is presented, detailing what data will be processed, for how long, and for what purpose (e.g., drowsiness detection, attention monitoring). The user may grant or reject consent, thereby exercising the GDPR rights to informed choice, transparency, and autonomy (Articles 6, 7, 21).

Once consent is granted, data flows to the Data Controller-typically an OEM or fleet operator-who determines the purpose and means of processing. The controller enforces privacy-by-design (PbD) and privacy-by-default (PbDf) principles, manages lawful bases for processing, and oversees Data Protection Impact Assessments (DPIAs) when large-scale monitoring or high-risk processing occurs. The controller also serves as the main point of contact for GDPR-based user requests, such as access, erasure, or rectification, ensuring that any user inquiry receives a clear and understandable response through the feedback mechanism depicted in the lower layer of the figure\ref{fig:Figure3}.

The controller may delegate certain operations to a Data Processor, such as a cloud or edge-based AI service. The processor operates solely under documented instructions, handling pseudonymized or anonymized data for tasks like model training or performance validation. Ethical safeguards include maintaining version logs, ensuring representativeness in AI training, and applying strong encryption (e.g., AES-256) for all data storage. When processing or storage occurs outside the EU, additional guarantees-such as Schrems II clauses or Standard Contractual Clauses (SCCs)-are applied to maintain GDPR equivalence.

The data storage node represents the final repository of processed information, secured via encryption and strict access control. A monitoring and vigilance layer continuously supervises the integrity of these operations, tracing data flows, auditing compliance, and activating remediation protocols in the event of misuse or unauthorized access.

The bottom tier of the framework-the feedback, monitoring, and vigilance modules-ensures that ethical governance remains dynamic rather than static. This layer embodies ongoing accountability, involving data protection officers (DPOs), ethics committees, and regulatory bodies who periodically review the fairness, proportionality, and social impact of DMS data practices.

Overall, this integrated framework establishes a closed ethical loop: user consent and rights flow upward to enable data processing, while transparent communication, oversight, and corrective mechanisms flow back downward to reinforce user trust. It demonstrates how GDPR principles-lawfulness, fairness, transparency, minimization, and accountability-can be operationalized within intelligent vehicle ecosystems.

\subsection{Stakeholder Roles and Responsibilities}
In the context of DMS, stakeholder roles must be carefully delineated to ensure transparency, accountability, and compliance with data protection laws such as the GDPR. The ethical implications of data collection, processing, and sharing are directly tied to how responsibilities and rights are distributed among key actors: data controllers, data processors, and data subjects. 
\subsubsection{Data Subjects: Rights and Autonomy}

Drivers and passengers whose biometric, behavioral, or contextual data is collected by DMS, are the primary data subjects and hold specific rights under the GDPR. These rights are essential to preserving their autonomy and agency. Transparency is critical to empowering individuals and ensuring their awareness of how and why their data is processed \citep{benyahya2022interface}. Key rights include:

\begin{itemize}
    \item The right to informed consent or to object to processing (Articles 6, 7, and 21).
    \item The right to access and understand what data is collected and how it informs automated decisions or interventions (Article 15).
    \item The right to data portability and erasure (Articles 20 and 17), particularly relevant in fleet or shared vehicle settings.
    \item The right not to be subject to decisions based solely on automated processing with significant legal or similar effects (Article 22), unless safeguards are explicitly in place.
\end{itemize}

However, the right to erasure is not absolute. It can be limited under GDPR Article 17(3), particularly where:

\begin{itemize}
    \item The data is required for compliance with a legal obligation (e.g., accident reconstruction, criminal investigation).
    \item The data is necessary for public interest or safety-critical purposes (e.g., ensuring road safety or maintaining system logs for liability).
    \item The data has been effectively anonymized and is no longer attributable to a data subject.
\end{itemize}

In such cases, the data controller must clearly communicate the rationale for denying or limiting an erasure request and provide appeal mechanisms or escalation pathways (e.g., DPO review or data protection authority oversight).

By embedding these safeguards, the framework supports a balanced interpretation of data ownership: one that respects user control and deletion rights while ensuring the functional and legal integrity of shared mobility and safety systems.

\subsubsection{Data Controllers: Role and Accountability}

In ethical frameworks, OEMs or fleet operators typically assume the role of data controllers, as defined under Article 4(7) of the GDPR. They determine the purposes and means of personal data processing - for example, analysing driver alertness or behavioural patterns via DMS. As controllers, they are responsible for:

\begin{itemize}
    \item Ensuring lawful processing bases (e.g., consent or legitimate interest) under Article 6.
    \item Implementing core data protection principles such as minimization, purpose limitation, and transparency (Articles 5, 13-14).
    \item Conducting and documenting Data Protection Impact Assessments (DPIAs) where required (Article 35), especially when processing involves systematic monitoring or poses high risks to individuals’ rights and freedoms.
    \item Facilitating GDPR rights such as access, rectification, objection, or erasure (Articles 15-22).
    \item Deciding whether data should be processed locally (e.g., edge processing within the vehicle) or via cloud infrastructure - and ensuring that if processing occurs outside the EU, a mandatory DPIA and appropriate safeguards (e.g., Standard Contractual Clauses (SCCs), encryption, anonymization) are in place.
\end{itemize}

\subsubsection{Data Processors: Operational Responsibility}

Third-party vendors, analytics providers, or cloud infrastructure services often serve as data processors, acting on behalf of the data controller. Under Article 28 of the GDPR, processors are required to:

\begin{itemize}
    \item Process personal data only on documented instructions from the controller.
    \item Implement appropriate technical and organizational safeguards, including pseudonymization, encryption, and access controls.
    \item Enter into Data Processing Agreements (DPAs) that clearly outline responsibilities, breach notification timelines, and compliance obligations.
    \item Cooperate with the controller in fulfilling data subject rights, conducting DPIAs, and responding to supervisory authorities.
    \item Ensure, in scenarios involving cross-border data transfers (particularly outside the EU), that hosting and processing activities align with GDPR Chapter V provisions and controller mandates.
\end{itemize}

\section{Proposed Framework Responses to Identified Ethical Challenges}

This section outlines how the proposed framework systematically addresses the ethical challenges associated with in-cabin sensing in DMS. For each identified ethical risk-ranging from privacy intrusion and data misuse to transparency, accountability, and user autonomy-the framework provides a structured and actionable response.

\begin{figure*}[htbp]
    \centering
    \includegraphics[width=0.99\textwidth]{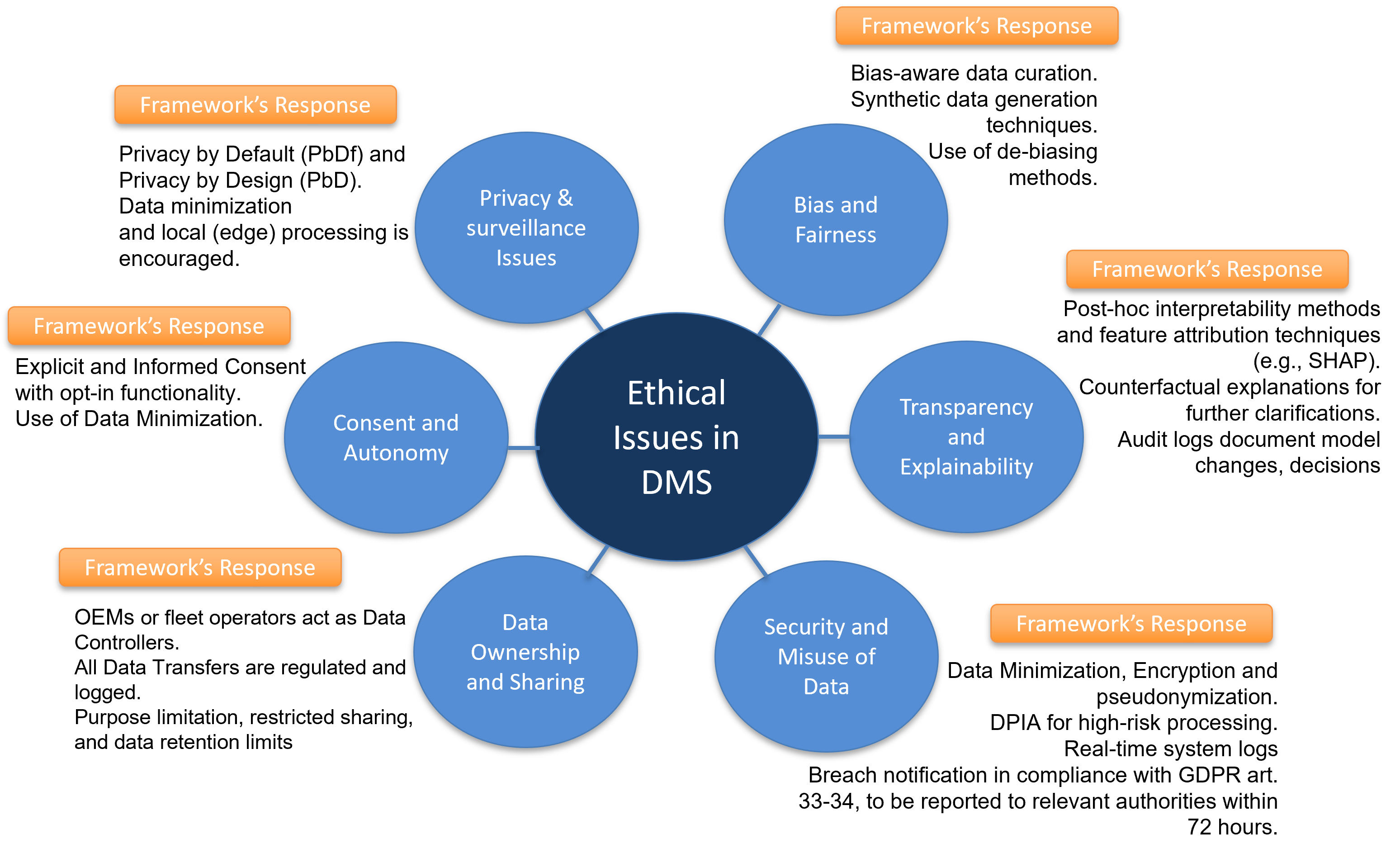}
    \caption{Overview of how the proposed framework addresses key ethical challenges in DMS in-cabin sensing.}
    \label{fig:Figure4}
\end{figure*}

Figure~\ref{fig:Figure4} presents an overview of how the framework responds to each ethical challenge identified in Section~\ref{sec:E-challenges}. Each response is discussed in two parts:

\begin{itemize}
\item \textbf{Design Response} – The technical or procedural safeguards integrated into the system architecture.
\item \textbf{Implementation Guidelines} – Practical recommendations for applying these safeguards in real-world deployment.
\end{itemize}

This structured approach ensures that ethical considerations are not treated as abstract concerns but are translated into actionable design and policy elements that can guide developers, regulators, and practitioners in deploying responsible and human-centric Driver Monitoring Systems.

\subsection{Addressing Privacy and Surveillance Concerns}

To mitigate the ethical challenges surrounding privacy and surveillance outlined in the previous section, our framework adopts a dynamic, process-oriented interpretation of PbD and PbDf, in alignment with GDPR Articles 25, 32, 35, and 36. Rather than relying on a static checklist of Privacy-Enhancing Technologies (PETs), PbD is treated as an evolving design philosophy that integrates both technical and organizational safeguards throughout the entire data lifecycle - from data collection to storage, processing, and potential transfer~\citep{ENISA2021-dddd}. 

In practical terms, the framework emphasizes the minimization of high-resolution video or audio input to protect physical privacy and favors less intrusive sensing modalities such as infrared, thermal, event-based cameras, or radar, where functionally adequate. Data collection is strictly limited to safety-critical purposes and governed by conservative default settings that reduce unnecessary capture. A dedicated Consent Request module transparently communicates the scope of AI-based processing, data retention periods, and the extent of data collection to users.

Edge (local) processing is prioritized to reduce reliance on centralized infrastructure, while strong anonymization and pseudonymization techniques - including context-aware anonymization in accordance with Opinion 05/2014 \citep{data2012article} - are employed where needed. When remote processing is necessary, it is secured using AES-256 encryption and Schrems II-compliant server infrastructure. Furthermore, Data Protection Impact Assessments (DPIAs) are integrated into the early design stages, particularly for high-risk deployments \citep{force2012data}, and the framework ensures ongoing accountability through continuous oversight by independent privacy advocates and ethics experts.

\subsection{Addressing Consent and Autonomy}

To address the ethical challenge of maintaining user autonomy in DMS, our framework operationalizes the principle of explicit, informed consent, as defined under Article 4(11) and governed by Article 5 of the GDPR. Given the continuous collection of highly sensitive behavioural data - such as gaze direction, facial expressions, and voice patterns - consent must be freely given, specific, informed, and unambiguous, with a clear affirmative action from the data subject and the ability to revoke consent at any time, in accordance with Article 7 \citep{regulation2018general}. In response, the framework implements a comprehensive consent management module designed to support not only initial consent capture but also an ongoing, user-centric model of consent.

This module enables dynamic and granular user control over multiple dimensions, including specific data streams (e.g., vision based, audio or radar sensor), the location of data processing (e.g., edge or cloud), and consent preferences. Modular opt-in pathways allow users to selectively engage with particular use cases \citep{bonnefon2016social}, while transparent communication strategies are employed to clarify trade-offs - for instance, reduced system functionality in exchange for limited data sharing - and to explain the minimum data necessary for core safety functions \citep{regulation2018general}. For example, in a multimodal DMS setup that includes camera, radar, and microphone inputs, a privacy-conscious driver may choose to opt out of camera-based sensing while retaining radar and audio modalities. In such a case, system functionality would be restricted to what can be inferred from the remaining modalities. This could enable limited features such as detecting drowsiness through snoring or abnormal breathing patterns via audio, or identifying the presence of a child in the backseat using radar-based motion and respiration sensing. However, more advanced capabilities such as precise gaze estimation, real-time facial expression analysis, or yawn detection would be unavailable due to the absence of visual input. By making such trade-offs transparent and explaining the safety-critical baseline functionality (e.g., distraction detection, child presence alerts), the framework supports informed user decisions while maintaining essential protective features.

Real-time dashboards or mobile applications allow users to review and withdraw consent easily \citep{bier2016privacyinsight}, and all use cases, third-party data sharing practices, and audit mechanisms are communicated clearly. Routine audits by Data Protection Officers (DPOs) ensure accountability, and the system is designed to comply with interoperable consent standards to maintain compatibility across platforms. Governance responsibilities are also clearly distributed: manufacturers are tasked with creating user-friendly consent interfaces, while accommodating the realities of distributed data responsibility in modern vehicle platforms.

 \subsection{Addressing Data Ownership and Sharing}

To address the complex and often contested issue of data ownership in DMS, our framework introduces clear governance mechanisms that reflect the multifaceted stakeholder environment of the automotive ecosystem. Ownership claims by OEMs, service providers, fleet operators, insurers, and end-users often intersect, creating ambiguity around who controls in-cabin data collection and use. This ambiguity intensifies when personal data is shared beyond the vehicle, necessitating compliance with GDPR requirements for lawful processing under Article 6 and, where purposes shift, additional justification under Article 6(4) \citep{kerber2018data} \citep{christensen2024pre}. Our framework resolves these tensions by assigning explicit data stewardship roles that distinguish between data controllers (typically OEMs or fleet operators, but sometimes drivers), data processors (such as DMS hardware vendors or third-party analytics providers), and data subjects (vehicle occupants) \citep{sargiotis2024data, doan2024framework}.

A central component is the Consent Request Module (detailed in Section 4.2.2), which facilitates informed and granular user consent while supporting the full exercise of data subject rights under Articles 15-21 of the GDPR, including access, rectification, objection, and erasure. To ensure accountability, all data flows involving third-party or off-vehicle processing are logged and monitored through auditable compliance modules. The framework also emphasizes data minimization and edge-based processing to reduce both privacy risk and ownership disputes. Cross-border data handling is governed by a stringent protocol: all remote processing defaults to EU-based servers, and international data transfers are permitted only after a Data Protection Impact Assessment (DPIA) is conducted by the data controller. Where such transfers are necessary, safeguards including Standard Contractual Clauses (SCCs), encryption, and strong anonymization techniques are applied. Together, these design choices uphold the principles of lawful, transparent, and controllable data sharing, while accommodating the realities of distributed data responsibility in modern vehicle platforms.

\subsection{Addressing Bias Mitigation and Fairness}

To prevent discriminatory outcomes and uphold principles of dignity, autonomy, and non-maleficence, our framework explicitly addresses the risk of algorithmic bias in DMS. Bias may arise from several sources, including skewed or incomplete training datasets, flaws in algorithmic development pipelines, or deployment in specific sociotechnical contexts that introduce unintended consequences \citep{barocas2016big}. Such biases can disproportionately affect individuals based on race, gender, age, or cognitive profiles, leading to unfair system behaviors and eroded user trust.

Our framework incorporates fairness as a foundational design principle across the entire development lifecycle. Bias-aware data curation is employed to ensure robust representation across age, gender, ethnicity, and cognitive traits in both training and evaluation datasets. To address underrepresentation - particularly for rare conditions (e.g., microsleep, panic, cognitive overload) or minority demographics (e.g., elderly, masked drivers, ethnic variance) - the framework leverages advanced generative AI tools. We use state-of-the-art generative adversarial networks (GANs) \citep{goodfellow2020generative} and latent diffusion models \citep{rombach2022high} to synthetically produce high-fidelity driver states, with fine-grained control over attributes such as facial expressions, occlusions (e.g., glasses, masks), and emotional conditions like stress or fatigue. These synthetic samples enable the augmentation of imbalanced datasets and support fairness auditing objectives such as demographic parity and group-wise recall balancing.

Generative AI is also applied to privacy-conscious modalities. For instance, we use identity-agnostic synthetically generated thermal LWIR \citep{farooq2025thermvision} and RGB face datasets \citep{farooq2023childgan}, \citep{farooq2025childdiffusion} to support GDPR-compliant training and augmentation pipelines, especially in low-light or night-time monitoring contexts. These modalities preserve essential features like head pose, eye closure, and facial temperature dynamics without requiring access to personally identifiable RGB images, thus mitigating both bias and privacy risks. Furthermore, generative face-to-video animation pipelines \citep{farooq2025synadult} enable the creation of realistic temporal sequences portraying gradual transitions such as alert-to-drowsy states, progressive yawns, gaze drift, and subtle cognitive distraction. Such synthetic video data bridges the gap between static-image-based training and real-time, sequential inference tasks needed for LSTM \citep{sherstinsky2020fundamentals} and Transformer-based driver state models \citep{huang2023gameformer}.

Beyond the technical pipeline, the framework advocates for participatory design involving historically marginalized stakeholders in the design, testing, and validation phases. This ensures that fairness is not solely approached algorithmically but also socio-technically, reinforcing ethical integrity across all stages of system deployment. Collectively, these generative AI-driven strategies not only enhance model robustness and generalization but also align with emerging regulatory standards such as ISO/PAS 21448 (Safety of the Intended Functionality) and anticipated EU AI Act compliance requirements.

\subsection{Transparency and Explainability}
In response to the ethical challenges identified earlier regarding transparency and explainability, the proposed framework adopts a multi-pronged strategy to address opacity in AI-powered DMS. These systems, particularly when deployed within AVs, often rely on complex and opaque models to process sensitive biometric and behavioral data. Such black-box approaches can compromise the user's ability to give meaningful consent, undermine their autonomy in human-machine interactions, and reduce the accountability of OEMs and service providers \citep{wachter2019right}. These concerns are not only ethical but also legal, as the GDPR Article 5 explicitly enshrines transparency as a core principle of lawful data processing, while Recital 39 further demands that personal data be communicated in “clear and plain language”~\citep{regulation2018general}.

To counter these issues, the framework emphasizes the creation of user-centric interfaces that explain system decisions and their implications in straightforward, accessible language. Rather than leaving users in the dark, these interfaces provide plain-language justifications for driver state assessments and subsequent system actions (e.g., alerting or disengagement). To further enhance interpretability, post-hoc explanation tools are integrated into the system. Techniques such as saliency maps (e.g., Grad-CAM) and feature attribution methods (e.g., SHAP) are used to visualize the model’s decision logic - highlighting, for instance, which visual regions (like drooping eyelids or gaze drift) were most influential in detecting drowsiness or distraction \citep{zhou2021predicting, Liu2024-wwww}.

Additionally, the framework leverages natural language explanations and counterfactual reasoning to help users understand model outcomes \citep{stepin2021survey}. For example, a system might convey that “If gaze had remained on-road for 20 more seconds, no alert would have been triggered,” offering intuitive insight into the model’s thresholds and behavior \citep{stepin2021survey}. Such transparency mechanisms are also designed to meet regulatory requirements, including GDPR’s obligation to provide meaningful information about the logic involved in automated decision-making \citep{ribeiro2016should}.

Taken together, these interventions ensure that DMS systems do not function as inscrutable black boxes but instead operate in a way that is interpretable, contestable, and aligned with ethical and legal standards. By making model behavior understandable to users and regulators alike, the framework enhances trust, supports user agency, and facilitates responsible deployment of AI in automated vehicles.

\subsection{Cybersecurity and Integrity Risks}
Cybersecurity threats in DMS include data breaches, spoofing of driver signals, and vulnerabilities introduced via third-party software or hardware components. These risks jeopardize both user privacy and system reliability, especially in safety-critical environments such as automated vehicles.

To address these concerns, the framework incorporates a multi-layered cybersecurity strategy designed to protect data integrity, system functionality, and user trust across the entire DMS lifecycle. At the core of this approach is the implementation of end-to-end encryption, ensuring that both raw and processed biometric data are securely transmitted and stored. All model updates and firmware patches are subject to secure boot protocols, digital signing, and hardware attestation to prevent tampering or unauthorized modifications.

Additionally, the system supports real-time integrity monitoring, using cryptographic checksums and anomaly detection algorithms to identify unexpected changes in model behavior or data flow. These are further reinforced by role-based access controls and secure identity management that restrict system-level permissions and log all access activities for traceability.

By combining robust technical safeguards with continuous threat assessment and compliance auditing, this solution aims to ensure that DMS deployments remain secure, reliable, and resilient in the face of evolving cyber threats.

\subsection{Psychological Impact}
Continuous driver monitoring can lead to discomfort, stress, or feelings of surveillance, particularly when users lack clarity on how the system operates or handles personal data.

To address this, the framework incorporates an iterative, stakeholder-driven feedback process aimed at enhancing user comfort and acceptance. Input from drivers, privacy advocates, and ethics experts is gathered through usability testing and surveys to assess perceived intrusiveness and emotional impact \citep{coyne2024understanding}.

These insights inform targeted design changes such as: 

\begin{itemize}
    \item Less intrusive sensing modalities (e.g., radar instead of RGB cameras),

    \item Configurable privacy settings, and
    
    \item Simplified explanations of how and why data is used.

\end{itemize}

By embedding psychological considerations into design and evaluation, the framework helps ensure DMS deployment supports user trust, well-being, and long-term acceptance.

\section{Risk Analysis and Failure Strategy \& Planning in Ethical DMS Design}
This section conducts a risk analysis based on the implementation of the proposed ethical framework, examining potential residual risks, failure modes, and the necessary mitigation strategies to ensure robust and trustworthy DMS deployment.
\subsection{Risk Analysis}

To operationalize the ethical framework within a DMS, we outline a robust strategy that transcends initial risk identification. This strategy focuses on mitigation, preparedness, and resilience, acknowledging the dynamic and high-stakes environment in which AVs operate.

\paragraph{Risk Mitigation}
\begin{itemize}
    \item DMS should limit data collection to essential parameters only (e.g., gaze direction, head pose), and favor local edge processing where feasible to reduce exposure.
    \item Encryption, pseudonymization, and strict access control should be implemented to protect sensitive biometric or behavioral data.
    \item Continuous validation for demographic fairness using diverse datasets and counterfactual analysis to prevent misclassification (e.g., drowsiness in older vs. younger drivers).
    \item Inclusion of user-facing explainability tools and regulatory access interfaces to pre-empt trust erosion and opacity issues.
\end{itemize}

\paragraph{Regular Review}
\begin{itemize}
    \item DPIAs should be re-triggered at each critical system update, especially when introducing new sensing modalities or machine learning (ML) models.
    \item Periodic reviews by interdisciplinary stakeholders (e.g., regulators, ethicists, user advocates) to ensure ongoing alignment with ethical and legal standards.
    \item Integration of driver-reported issues and system logs to refine training datasets and mitigate long-term drift in algorithm behavior.
\end{itemize}

\subsection{Incident Management and Failure Strategy}

When DMS errors occur-especially those with safety or ethical implications-a clearly structured incident response plan is essential.

\paragraph{Incident Response Plan}
\begin{itemize}
    \item Real-time system logs should capture input/output signals, driver state classification, and intervention triggers, stored in a privacy-respecting and secure format (e.g., edge-local with pseudonymization).
    \item A triaged event severity model should be used to determine escalation procedures - ranging from software bug fixes to full regulatory notifications.
\end{itemize}

\paragraph{Breach Notification}
\begin{itemize}
    \item If a personal data breach occurs, notifications should comply with GDPR Articles 33 and 34, informing relevant supervisory authorities and affected users within 72 hours.
    \item For AV-specific scenarios, breaches involving biometric or real-time behavioral profiling should trigger automated alerts to both OEMs and relevant authorities.
\end{itemize}

\paragraph{Post-Incident Review}
\begin{itemize}
    \item All high-severity incidents must trigger a post-incident review board (including technical, legal, and ethical roles).
    \item Reports should document: system state, failure root cause, user/system interaction logs, and remediation actions.
    \item Lessons learned must feed into training set updates, model auditing, and interface redesign if needed.
\end{itemize}

\subsection{Accountability and Governance}

Establishing clear roles and responsibilities is fundamental for ensuring both compliance and ethical accountability.

\paragraph{Data Protection Officer (DPO)}
\begin{itemize}
    \item The DPO oversees GDPR compliance across all DMS components-particularly regarding profiling, consent, and high-risk processing (e.g., fatigue detection).
    \item Acts as the primary liaison with supervisory authorities and ensures timely execution of DPIAs and breach notifications.
\end{itemize}

\paragraph{Audit and Compliance}
\begin{itemize}
    \item Periodic external audits (technical and ethical) are conducted to verify compliance with GDPR and emerging AI regulations (e.g., EU AI Act).
    \item Audit trails are automatically generated by the DMS, documenting access to personal data, decision making logic, and system interventions.
\end{itemize}

\paragraph{Incident Governance}
\begin{itemize}
    \item Governance committees internal or joint with regulators-must assess patterns of failures (e.g., demographic bias in detection).
    \item Formalize accountability through corrective actions, system level patches, and, if necessary, suspension of features pending reassessment.
\end{itemize}

\medskip

This dual approach proactive risk mapping and responsive failure strategy ensures that the ethical framework is not merely preventive, but adaptive and resilient. It reinforces user trust, regulatory compliance, and system integrity across evolving AV environments.

\subsection{Sharing Data with Third Parties }

In some cases, it may be necessary for either the car manufacturer or the car owner to share the footage from the cameras with third parties or the car manufacturer e.g. to improve the technical installations, for the car owner e.g. to assert legal claims following damage to the car. Such sharing with third parties is considered in the GDPR as a separate processing operation that requires its own processing basis in Art. 6 as well as compliance with Art. 6(4) if the data is disclosed for processing for a purpose other than that for which it was originally collected \citep{regulation2018general}. It follows from GDPR preamble recital 50 that "Indicating possible criminal acts or threats to public security by the controller and transmitting the relevant personal data in individual cases or in several cases relating to the same criminal act or threats to public security to a competent authority should be regarded as being in the legitimate interest pursued by the controller" \citep{regulation2018general}. This means that if the information from the cameras is forwarded to the police in order for the police to solve a crime, the controller has a legitimate interest in this and can find the basis for processing in Article 6(1)(f). However, here again, a balance must be struck as to whether the data subject’s interests take precedence over the legitimate interest of the controller \citep{regulation2018general}.  

But as a starting point, the data controller, including both the car manufacturer and the car owner, can send recordings from the cameras to the police if the recordings relate to a criminal offence. In Denmark, for instance, this has been the situation in a case where a young police officer was hit by a car while on duty, and the prosecution presented a video from a Tesla that filmed the collision as part of the evidence. In addition to the fact that the data controller can choose to provide the police with information from the cameras, the police can also obtain a warrant to obtain the footage from the cameras at the car owner’s premises. In that case, the processing will be based on a legal obligation incumbent on the controller, cf. GDPR Art. 6(1)(c) \citep{regulation2018general}. In addition to the car owner sharing information with the police, there have also been examples of car owners sharing footage from the camera on social media, such as Facebook, Instagram or YouTube. Such sharing would need to have a legal basis in the GDPR. If the owner chooses to share information on social media, the owner must therefore first and foremost assess whether it is "ordinary" personal data or sensitive personal data, and thus whether the basis for processing must be found in Art. 6 and/or Art. 9. It will probably in very few cases be legal for the car owner to share the recordings from the car if they contain sensitive data, unless the car owner has obtained consent to do so. For general personal data, the owner will most likely find a legal basis for processing in the form of consent (GDPR art. 6(1)(a) or the balancing of interests rule (GDPR art. 6(1)(f)). If the car owner shares, with a legal basis in GDPR art. 6(1)(f), the car owner must assess the owner’s interest in sharing images against the data subjects’ right to privacy. Again, the balance will depend on the car owner’s purpose for sharing, but there are likely to be compelling reasons for the car owner to share the footage \citep{regulation2018general}.

\vspace{1em}
\section{Comparison: Minimal Compliance vs Ethical Enhancements}

This section discusses the implementation requirements and associated burdens of the proposed ethical framework for DMS/ICS. To provide a structured comparison, Table \ref{tab:compliance_framework} contrasts three key dimensions for each ethical challenge:

\begin{itemize}
    \item \textbf{Minimal legal compliance (baseline)}: Practices commonly adopted to meet regulatory requirements.
    \item \textbf{Additional measures under the ethical framework}: Enhanced practices aimed at improving privacy, fairness, safety, and user acceptance.
    \item \textbf{Incremental burden}: Examples of the technical, operational, and financial overhead introduced by implementing these advanced measures.
\end{itemize}

\begin{table}
\centering
\scriptsize
\renewcommand{\arraystretch}{1.25}
\setlength{\extrarowheight}{2pt} % add a bit of vertical padding for readability
\begin{tabularx}{\textwidth}{
>{\raggedright\arraybackslash}p{3cm} % Challenge fixed width
>{\raggedright\arraybackslash}X % Minimal Legal Compliance flexible width
>{\raggedright\arraybackslash}X % Ethical Framework flexible width
>{\raggedright\arraybackslash}X % Incremental Burden flexible width
}
\toprule
\textbf{Challenge} & 
\textbf{Minimal Legal Compliance (Baseline)} & 
\textbf{Extra under the Ethical Framework} & 
\textbf{Incremental Burden (Examples)} \\
\midrule

Privacy-by-Design &
Dual-facing AI dash-cam uploads short clips; relies on legitimate interest. Basic TLS link and at-rest AES. &
Full on-device analytics; strong anonymisation; Data Protection Impact Assessment (DPIA) and privacy oversight. &
Powerful edge SoC; DPIAs, audits, privacy board. \\
\midrule

Consent \& User Control &
Driver notice in handbook and checkbox; no granular opt-out. &
Granular consent centre, privacy modes, withdrawal workflows. &
UI/backend/policy API development; Data Protection Officer (DPO) and legal time. \\
\midrule

Data Governance \& Cross-Border &
Single US cloud; covered by privacy policy. &
EU servers or Standard Contractual Clauses (SCCs); transfer logs; access reviews. &
EU cluster fees; log governance. \\
\midrule

Multi-sensor Safety (Radar/Thermal) &
IR camera only. &
Radar for vitals/child detection; thermal for fever/fatigue monitoring. &
+\$35–50 per radar; +\$200 per thermal sensor; extra install time; software fusion. \\
\midrule

Edge-efficiency vs Bandwidth &
Continuous upload; \$30/month data plan. &
Edge filtering; improved storage; Over-The-Air (OTA) infrastructure. &
Larger eMMC/SSD; secure OTA pipeline. \\
\midrule

Algorithmic Fairness \& Bias Audits &
Train on available footage; no fairness testing. &
Curated datasets; fairness metrics; remediation cycles. &
Long-tail data curation; ethics team; audit processes. \\
\midrule

Explainability \& Transparency &
Black-box machine learning; generic alerts. &
Human-readable explanations; regulator-grade logs. &
UI overlays; SHAP tools; long-term log storage. \\
\midrule

Security \& Incident Response &
Standard encryption; ad-hoc patches. &
Penetration tests; 72-hour breach playbook; audit trail. &
Annual contracts; on-call security staff. \\
\midrule

Health Intervention / Safe-stop &
Visual-based drowsiness detection. &
Radar-based vitals monitoring with auto-stop logic. &
Sensor costs; fail-safe certification; insurance. \\
\midrule

Psychological Well-being \& Acceptance &
No user involvement; mandatory fit. &
Workshops, pilots, privacy shutters, alert tuning. &
HR training; human–machine interface (HMI) iteration. \\
\midrule

Lifecycle Governance &
Internal Project Manager oversight. &
Cross-functional board; transparency reports. &
Compliance/ethics Full-Time Equivalent (FTE); audit costs. \\
\bottomrule

\end{tabularx}
\caption{Comparison between Minimal Compliance, Ethical Framework Enhancements, and Resulting Incremental Burdens}
\label{tab:compliance_framework}
\end{table}

Table~\ref{tab:compliance_framework} illustrates that while baseline compliance is often limited to minimal security measures, single-cloud data hosting, and simple consent mechanisms, this ethical framework advocates for privacy-preserving architectures (e.g., on-device analytics), multi-sensor safety integration (e.g., radar and thermal sensing), algorithmic fairness audits, explainability mechanisms, and stakeholder-centric design. These upgrades require additional resources, including higher-capacity hardware , e.g., edge system on chip (SoCs), increased engineering effort, e.g., user interface (UI) and policy application programming interface (APIs), and recurring operational costs (e.g., privacy audits, fairness evaluations, and insurance for safe-stop features).

While the incremental burdens - such as capital expenditure on sensors, extended development timelines, and additional compliance staffing - may initially appear significant, they are outweighed by the long-term benefits: enhanced user trust, regulatory readiness, improved system safety, and reduced reputational or legal risks.

The intention behind presenting this comparative table is to demonstrate that acting ethically in business may, in fact, be beneficial - not only from a societal or compliance perspective but also as a competitive differentiator.

In addition to ethical considerations, hardware implications must also be addressed. Table~\ref{tab:bom_comparison} compares the per-vehicle hardware costs of a standard dual-camera dash-cam system with those of an ethically enhanced configuration. The latter incorporates radar sensing, higher-capacity edge compute modules, and secure storage to meet the requirements of the proposed framework.

As shown, upgrading from a baseline DMS to the ethical specification results in a modest per-vehicle hardware increase of approximately \$80–150, primarily due to the added radar, enhanced compute, and secure storage. Installation overhead contributes an additional \$50–75 per vehicle \citep{icesystems2025_commercial_alarm}.

However, the dominant cost driver lies in ongoing operational commitments: staffing of DPOs, conducting annual privacy and fairness audits, running DPIAs, and maintaining secure infrastructure. For a fleet of 1,000 vehicles, these recurring governance costs can reach mid-six figures annually.

Despite this, such investments yield tangible benefits:
\begin{itemize}
    \item Regulatory preparedness for evolving mandates under the EU AI Act, GDPR, and ISO standards.
    \item Increased user trust and driver acceptance, especially in safety-critical and privacy-sensitive domains.
    \item Long-term cost offsets via reduced bandwidth usage (due to edge processing) and minimized legal liability.
\end{itemize}

Importantly, the shift to on-device analytics-while initially resource-intensive-can reduce recurring data transmission and cloud storage expenses. Over a multi-year horizon, this leads to a favourable Total Cost of Ownership (TCO), particularly when reputational and legal risks are accounted for.

\begin{table}
\centering
\scriptsize
\renewcommand{\arraystretch}{1.25}
\setlength{\extrarowheight}{2pt}
\begin{tabularx}{\textwidth}{
>{\raggedright\arraybackslash}p{5cm}
>{\raggedright\arraybackslash}X
}
\toprule
\textbf{Config} & \textbf{Typical BOM} \\
\midrule

Baseline dual-cam dash-cam & 
\$300 – \$600 (camera + compute + LTE) \\

\midrule

Ethical spec (add radar + larger SoC + extra storage \& encryption chip) & 
\$380 – \$750 (+\$35–50 radar + \$20–40 compute uplift + \$10–20 storage/encryption) \\

\bottomrule
\end{tabularx}
\caption{Estimated Bill of Materials (BOM) Comparison Between Baseline and Ethical DMS Configurations}
\label{tab:bom_comparison}
\end{table}

\section{Conclusion}

% \subsection{Ethical challenges and framework responses}

This paper argues that DMS are essential safety technologies for autonomous and semi-autonomous vehicles, especially at SAE Levels 2-4 where human oversight remains necessary. DMS use cameras, biometrics, radar, and thermal sensing to detect fatigue, distraction, and health emergencies. While improving safety, continuous monitoring raises significant ethical concerns around privacy, autonomy, data ownership, bias, and psychological impact.

This paper suggests that DMS ethics should not be subsumed under general AV ethics. Instead, it proposed a layered interface framework spanning the AV-DMS, DMS-driver, and AV-driver relationships, with emphasis on the DMS-driver interface where concerns are most immediate and tied to sensitive biometric data.

This paper reviews regulatory baselines - GDPR, the EU AI Act, and IEEE Ethically Aligned Design - noting that GDPR strongly protects privacy yet can slow innovation, the AI Act introduces risk-based duties with some overlap, and IEEE EAD promotes anticipatory ethics without enforcement.

Key ethical challenges highlighted include:
\begin{itemize}
    \item Privacy \& surveillance (panoptic effect from in-cabin sensing)
    \item Consent \& autonomy (bundled/implicit consent, limited agency)
    \item Data ownership \& sharing (ambiguity across drivers, OEMs, vendors)
    \item Bias \& fairness (unequal performance across demographics)
    \item Transparency \& explainability (black-box models)
    \item Security risks (breach/misuse of biometric data)
    \item Psychological burden (stress from constant monitoring)

\end{itemize}

To address these, this paper argues for a principle-based, modular, and adaptable ethical framework grounded in transparency, accountability, fairness, and autonomy. It promotes privacy-by-design/default, edge processing to minimize data transfer, dynamic granular consent, bias-aware dataset curation with ongoing audits, explainability tooling, and independent oversight-treating ethics as a continuous lifecycle, not a one-off checklist.

The paper further suggests that moving beyond minimal legal compliance entails costs-additional sensors (e.g., radar/thermal), stronger edge compute, and recurring governance (DPOs, DPIAs, audits). However, it argues these are strategic investments that enhance trust, regulatory readiness, safety, and reputational resilience, making ethical design a competitive advantage.

% -------------------------------------------
% BIBLIOGRAPHY
% -------------------------------------------
\bibliographystyle{elsarticle-harv} % TR Part F is Harvard/author-year style
\bibliography{references}

\end{document}